\RequirePackage{fix-cm}
\documentclass[smallextended]{svjour3}       
\smartqed  
\usepackage{graphicx}
\usepackage{dsfont}
\usepackage{amsmath}
\usepackage{amssymb}
\usepackage[utf8]{inputenc}
\usepackage{multirow}
\usepackage{xcolor}
\usepackage{epsfig}
\usepackage[figuresright]{rotating}
\usepackage{subfig}
\usepackage{array,booktabs}
\usepackage{hyperref}
\usepackage{natbib}
\usepackage{graphicx}
\pdfmapfile{+txfonts.map}
\begin{document}

\title{Composite distributions in the social sciences:
A comparative empirical study of firms' sales distribution for France, Germany, Italy, Japan, South Korea, and Spain\thanks{This study was supported by JESS KAKENHI Grant Numbers 17K01277, 21K04557, 18H05217, by \emph{Deutsche Forschungsgemeinschaft (DFG)} Grant Number 455257011, PID2020-112773GB-I00 of the Spanish \emph{Ministerio de Ciencia e Innovaci\'on} and ADETRE Reference Group S39\_20R of \emph{Gobierno de Arag\'on.}}
}


\titlerunning{Composite distributions for firm sizes}        

\author{Arturo Ramos\and
Till Massing\and
Atushi Ishikawa\and
Shouji Fujimoto\and
Takayuki Mizuno
}

\authorrunning{Ramos et al.} 

\institute{A. Ramos\at
Departmento de An\'alisis Econ\'omico\\
Universidad de Zaragoza, Zaragoza, Spain\\
\email{aramos@unizar.es}
\and
T. Massing \at
                     Universit\"{a}t Duisburg-Essen,
 Fakult\"{a}t f\"{u}r Wirtschaftswissenschaften\\
 Lehrstuhl für \"{O}konometrie, Essen, Germany  \\
                        \email{till.massing@uni-due.de}
                        \and
           A. Ishikawa \at
           Department of Economic Informatics\\
            Kanazawa Gakuin University, Kanazawa, Japan\\
            \email{ishikawa@kanazawa-gu.ac.jp}
\and
S. Fujimoto \at
Department of Economic Informatics\\
Kanazawa Gakuin University, Kanazawa, Japan\\
              \email{sjfjmt@gmail.com}           \and
                    T. Mizuno \at
            National Institute of Informatics, Tokyo, Japan\\
The Graduate University for Advanced Studies [SOKENDAI], Tokyo, Japan\\
Center for Advanced Research in Finance, The University of Tokyo, Tokyo, Japan\\
\email{mizuno@nii.ac.jp}
}

\date{Received: date / Accepted: date}

\maketitle

\begin{abstract}
We study 17 different statistical distributions for sizes obtained {}from the classical and recent literature to describe a relevant variable in the social sciences and Economics, namely the firms' sales distribution in six countries over an ample period. We find that the best results are obtained with mixtures of lognormal (LN), loglogistic (LL), and log Student's $t$ (LSt) distributions.
The single lognormal, in turn, is strongly not selected. We then find that the whole
firm size distribution is better described by a mixture, and there exist subgroups of firms. Depending on the method of measurement, the best fitting distribution cannot
be defined by a single one, but as a mixture of at least three distributions or even four or five.
We assess a full sample analysis, an in-sample and out-of-sample analysis, and a doubly truncated sample analysis.
We also provide the formulation of the preferred models as solutions of the Fokker--Planck or forward Kolmogorov equation.
\end{abstract}

\section{Introduction}

The study of the size variables in the social sciences and Economics has a long tradition like the use of a power-law in the study of the income distribution \citep{Par1896}, the application to the upper tail of the size distribution of cities \citep{Sin36,zipf1949human}, see e.g., also the recent article by \cite{Par21}. On its side, the distribution of firms has been the subject of early studies as the classic one by R. Gibrat \citep{Gib31}, where the lognormal distribution has been proposed, based on the now called Gibrat's law or the law of proportionate effect.
Classical references have proposed the lognormal distribution for firm size and variations \citep{SimBon58,Qua66,Cla79,CabMat03}.
In the city size distribution, the lognormal has been proposed, e.g., by \cite{Par85} and \cite{eeckhout2004gsl}.

More recent publications dealing with firm sizes' distribution and related matters could be, for example, \cite{Ree01,Ree02,Ree03,reed2004double,
GioLevRan11,GuoXuCheWan13,IoaSko13,Tod17,CorMorPer17,KwoNad19,BanChiPrePueRam19,
Su19,GuaTos18,GuaTos19,GuaTos19b,PueRamSan20,PueRamSanArr20,
CamRam21,Reg21,Ish21}.
Recent articles on the issue about whether the upper tail is lognormal or Pareto could be, e.g., \cite{Per05,ClaShaNew09,BeeRicSch11,BeeRicSch13,Bee15,BeeRicSch17,ChuDicNad19,SchTre19,Bee22,ChaFla22}.
Also, there are studies aiming at the explanation of economically important quantities by mixtures of distributions \citep{BelCleBul14,Kun20}, being quantities for which sub-populations of it may exist \citep{McLPee00}.
Let us mention as well that this research has roots on previous one of several years ago \citep{GonRamSan14,Pue15,RamSan15,Ram17}.

For the theoretical explanations of the so-called Zipf's law and Gibrat's law standard references are \cite{Gab99,Gab09} and \cite{Gib31}, and in modern terms, they can be understood, respectively, as distributions that are solutions to Fokker--Planck or forward Kolmogorov equations associated to It\^{o} differential equations \citep{Ord74,Gar04,ItoMcK96,Kyp06} with, respectively, reflective lower barrier \citep{Har85} or not, but both with constant diffusion and drift terms. Let us remark that the solutions to Fokker--Planck or forward Kolmogorov equations are mathematical concepts, but have been accepted by the Economics community after \cite{Gab99,Gab09} as plausible economical explanations of the observed empirical laws, as those papers are widely cited afterwards.

However, there is a more recent reference on the subject that sheds light on a fact known after Dupire \citep{Dup93,Dup94} but somehow overlooked in the literature that many probability density functions can be obtained in that way.
In fact, following \cite{PenPueRamSan22}, let
$y=\ln(x)$ be the logarithm of the firm's sales variable $x>0$. We assume that its evolution or dynamics is governed by the It\^{o} differential equation (see, e.g., \cite{Ord74,Gar04})
\begin{equation*}
dy_{t}=b(y_{t},t)dt+\sqrt{a(y_{t},t)}dB_{t}  
\end{equation*}%
where $B_{t}$ is a standard Brownian motion (Wiener process) (see, e.g., \cite{ItoMcK96,Kyp06} and references therein). The quantity $a(y_{t},t)$
corresponds to the \emph{diffusion term}, and $b(y_{t},t)$ to the \emph{drift term}. This process can be associated to the \emph{forward Kolmogorov
equation} or \emph{Fokker-Planck equation} for the time-dependent
probability density function (PDF) (conditional on the initial data) $f(y,t)$ (see
also \cite{Gab99,Gab09}):
\begin{equation}
\frac{\partial f(y,t)}{\partial t}=-\frac{\partial }{\partial y}%
(b(y,t)f(y,t))+\frac{1}{2}\frac{\partial ^{2}}{\partial y^{2}}(a(y,t)f(y,t))
\label{fpeq}
\end{equation}

Since the probability density function $f(y,t)$ is evolving on time and perhaps there is no limiting stationary distribution, let us propose a way of solving the Fokker--Planck equation for the mentioned $f(y,t)$ by specifying the diffusion term and the drift term (see, e.g., \cite{Otu21} for another recent approach to time-dependent solutions of the Fokker--Planck equation). In fact, if we take $a(g,t)=s^2$, where $s>0$ is a real constant, then by choosing
\begin{equation}
b(y,t)=\frac{s^2}{2 f(y,t)}\frac{\partial f(y,t)}{\partial y}-\frac{1}{f(y,t)}
\frac{\partial\,{\rm cdf}(y,t)}{\partial t}\,,
\label{solb}
\end{equation}
where ${\rm cdf}(y,t)$ is the cumulative distribution function (CDF) corresponding to $f(y,t)$, this solves the Fokker--Planck equation for $f(y,t)$ \citep{Dup93,Dup94}. However, we remark that we do not claim that such a solution with this choice of $a(y,t)$ and $b(y,t)$ is unique, only that $f(y,t)$ is a solution by construction. Also, $b(y,t)$ might have bounded discontinuities in the variable $y$, in a finite number of points in the domain \citep{GikSko07}. And a third remark is that we may add to the expression of $b(y,t)$ above a term of the form $h(t)/f(y,t)$, where $h(t)$ is an arbitrary function of $t$.

Thus we intend to fill the gap from theory (that many PDFs can be generated in this way) to empirical facts (what probability density function or functions is or are observed empirically), employing this empirical work, for the manufacturing firm's sales distribution for six countries in an ample period.

We offer a panoramic view of the subject for manufacturing firm's sales distribution since we have at hand ample databases for several countries, covering the full sales distribution, and not only the upper tail.
For that, we select some distributions that have been proposed recently for size studies, including firms' sizes, that satisfy that the PDFs are differentiable at least three times in the PDF parameters, and that have a CDF expressible using elementary or special functions known to date.

The rest of the paper is organized as follows. Section~\ref{distros} describes the distributions to be compared. Section~\ref{data} describes the databases used. Section~\ref{results} offers the results. Finally, we end with some conclusions.

\section{The distributions}\label{distros}

As mentioned earlier, $x>0$ will denote the sales of the firm in question.
The first distribution that we will consider, as a baseline model, is the usual lognormal distribution (LN), given by
\begin{equation}
f_{{\rm LN}}(x;\mu,\sigma)=\frac{1}{\sigma\sqrt{2\pi}x}
\exp{\left(-\frac{(\ln(x)-\mu)^2}{2\sigma^2}\right)}\,,
\label{modelLN}
\end{equation}
where $\mu\in\mathbb{R}$, $\sigma>0$.
The corresponding CDF is
\begin{equation*}
\Phi(\ln(x);\mu,\sigma)
=\frac{1}{2}+\frac{1}{2}{\rm erf}\left(\frac{\ln(x)-\mu}{\sigma\sqrt{2}}\right)\,.
\end{equation*}
where ${\rm erf}$ denotes the error function associated to the standard normal distribution.

The second distribution in our study is the double Pareto lognormal (DPLN) \cite{Ree02,Ree03,reed2004double,Man09,giesen2010size},
given by
\begin{eqnarray}
& &f_{\rm DPLN}(x;\alpha,\beta,\mu,\sigma)\nonumber\\
& &=\frac{\alpha\beta}{(\alpha+\beta)x}
\left(\exp\left(-\alpha(\ln(x)-\mu)+\frac{1}{2}\alpha^2\sigma^2\right)
\Phi(\ln(x);\mu+\alpha \sigma^2,\sigma)\right.\nonumber\\
& &\quad\quad\quad\left.+\exp\left(\beta(\ln(x)-\mu)+\frac{1}{2}\beta^2\sigma^2\right)
\Phi(-\ln(x);-\mu+\beta\sigma^2,\sigma)\right)
\label{modelDPLN}
\end{eqnarray}
where $\mu\in\mathbb{R}$, $\alpha,\beta,\sigma>0$ are the four parameters of the distribution. It has the property that it approximates different power-laws in each of its two tails: $f_{\rm DPLN}(x)\approx x^{-\alpha}$ when $x\to\infty$ and
$f_{\rm DPLN}(x)\approx x^{\beta}$ when $x\to 0$. The body is approximately lognormal, although it is not possible to delineate exactly the switch between the lognormal and the power-law behaviors since the DPLN distribution is the log version of the convolution of an asymmetric double
Laplace with a normal distribution.

The third distribution in our study is the Generalized Beta of the second kind (GB2) \citep{McD84,McDXu95,KleKot03} and
depends on four parameters with probability density function:
\begin{equation}
f_{{\rm GB2}}(x;a,b,p,q)=\frac{a x^{a p-1}}{b^{a p}B(p,q)\left(1+\left(x/b\right)^a\right)^{p+q}}
\label{modelGB2}
\end{equation}
where $B(p,q)=\int_0^1 t^{p-1}(1-t)^{q-1}\,dt$ is the Beta function, and $a,b,p,q>0$ are the four distribution parameters.

A way of modelling firm size distribution with semi-nonparametric densities is found in \cite{CorMorPer17}. The log-semi-nonparametric (LNSNP) density is defined using the Hermite polynomials until degree 4 and the lognormal density as follows:
\begin{eqnarray}
& &f_{\rm LNSNP}(x;\mu,\sigma,d_1,d_2,d_3,d_4)\nonumber\\
& &=f_{{\rm LN}}(x;\mu,\sigma)
(1+d_1 h_1(z)+d_2 h_2(z)+d_3 h_3(z)+d_4 h_4(z))
\label{modelLNSNP}
\end{eqnarray}
where $\mu\in\mathbb{R}$, $\sigma>0$, $d_1,d_2,d_3,d_4\in\mathbb{R}$ are the distribution parameters, $z$ is a shorthand for ${\displaystyle \frac{\ln(x)-\mu}{\sigma}}$
and
\begin{eqnarray}
&& h_1(z)=z\nonumber\\
&& h_2(z)=z^2-1\nonumber\\
&& h_3(z)=z^3-3z\nonumber\\
&& h_4(z)=z^4-6z^2+3\nonumber
\end{eqnarray}
are the cited Hermite polynomials.

The $\ell$-mixtures of lognormal distributions ($\ell$LN) for $\ell\geq 2$ are defined, for example, as follows \citep{McLPee00,GuaTos18,GuaTos19b,GuaTos19,KwoNad19,Su19,
PueRamSan20,PueRamSanArr20,CamRam21}
\begin{eqnarray}
f_{{\rm \ell LN}}
(x;\mu_1,\sigma_1,\dots,\mu_\ell,\sigma_\ell,p_1,\dots,   p_{\ell-1})
&=&\sum_{j=1}^{\ell-1} p_j f_{{\rm LN}}(x;\mu_j,\sigma_j)\nonumber\\
& &+\left(1-\sum_{j=1}^{\ell-1}p_j\right)f_{{\rm LN}}(x;\mu_\ell,\sigma_\ell)\nonumber
\end{eqnarray}
where $0\leq p_1,\dots,p_{\ell-1}\leq 1$, $0\leq 1-\sum_{j=1}^{\ell-1}p_j\leq 1$, and $\mu_i\in\mathbb{R}$, $\sigma_i>0$, $i=1,\dots,\ell$.

The specific lognormal mixtures that we will consider are
\begin{eqnarray}
&& \ell=2: {\rm 2LN} \label{model2LN}\\
&& \ell=3: {\rm 3LN} \label{model3LN}\\
&& \ell=4: {\rm 4LN} \label{model4LN}\\
&& \ell=5: {\rm 5LN} \label{model5LN}
\end{eqnarray}

Likewise, the loglogistic (LL) distribution has the probability density function
$$
f_{{\rm LL}}(x;\mu,\sigma)
=\frac{\exp{\left(-\frac{\ln(x)-\mu}{\sigma}\right)}}
{x\sigma
\left(1+\exp{\left(-\frac{\ln(x)-\mu}
{\sigma}\right)}\right)^2}
$$
where $\mu\in\mathbb{R},\sigma>0$ and $x>0$.
The $\ell$-mixtures of loglogistic distributions ($\ell$LL) for $\ell\geq 2$ can be defined, analogously, as \citep{McLPee00,PueRamSan20}
\begin{eqnarray}
f_{{\rm \ell LL}}
(x;\mu_1,\sigma_1,\dots,\mu_\ell,\sigma_\ell,p_1,\dots,   p_{\ell-1})
&=&\sum_{j=1}^{\ell-1} p_j f_{{\rm LL}}(x;\mu_j,\sigma_j)\nonumber\\
& &+\left(1-\sum_{j=1}^{\ell-1}p_j\right)f_{{\rm LL}}(x;\mu_\ell,\sigma_\ell)\nonumber
\end{eqnarray}
where $0\leq p_1,\dots,p_{\ell-1}\leq 1$, $0\leq 1-\sum_{j=1}^{\ell-1}p_j\leq 1$, and $\mu_i\in\mathbb{R}$, $\sigma_i>0$, $i=1,\dots,\ell$.

The specific loglogistic mixtures that we will consider are
\begin{eqnarray}
&& \ell=2: {\rm 2LL} \label{model2LL}\\
&& \ell=3: {\rm 3LL} \label{model3LL}\\
&& \ell=4: {\rm 4LL} \label{model4LL}\\
&& \ell=5: {\rm 5LL} \label{model5LL}
\end{eqnarray}

Some other mixtures could be considered. We adapt three other mixtures that appeared firstly in
\cite{MasPueRam20} and afterwards in \cite{MasRam21}, called 2St12, 2St39, and 3St, to yield the 2LSt12, 2LSt39, and 3LSt, and other supplementary two that will be called 4LSt, 5LSt.
They are as follows.
If the log version (to correctly deal with a size variable as it is firms' sales) of the non-stan\-dar\-di\-zed Student's $t$ distribution is \citep{JohKotBal95}
\begin{eqnarray}
f_{{\rm LSt}}(x;\mu,\sigma,\nu)=
\frac{\Gamma\left(\frac{\nu+1}{2}
\right)}{x\Gamma\left(\frac{\nu}{2}\right)\sqrt{\pi \nu}\sigma}
\left(1+\frac{1}{\nu}\left(\frac{\ln(x)-\mu}{\sigma}
\right)^2
\right)^{-\frac{\nu+1}{2}}\nonumber
\end{eqnarray}
where $\mu\in\mathbb{R},\sigma>0$ and $\nu>0$ is the number of degrees of freedom parameter, and $\Gamma(\cdot)$ denotes the Gamma function. This distribution, without logs,
has been used to study the log-growth rates of the size of German cities \citep{SchTre16}.

The $\ell$-mixtures of LSt distributions for $\ell\geq 2$ can be defined, analogously, as
\citep{McLPee00}
\begin{eqnarray}
& &f_{{\rm \ell LStudent}}
(x;\mu_1,\sigma_1,\nu_1,\dots,\mu_\ell,\sigma_\ell,\nu_\ell,p_1,\dots,   p_{\ell-1})\nonumber\\
& &=\sum_{j=1}^{\ell-1} p_j f_{{\rm LSt}}(x;\mu_j,\sigma_j,\nu_j)\nonumber\\
& &+\left(1-\sum_{j=1}^{\ell-1}p_j\right)f_{{\rm LSt}}(x;\mu_\ell,\sigma_\ell,\nu_\ell)\nonumber
\end{eqnarray}
where $0\leq p_1,\dots,p_{\ell-1}\leq 1$, $0\leq 1-\sum_{j=1}^{\ell-1}p_j\leq 1$, and $\mu_i\in\mathbb{R}$, $\sigma_i>0$, $\nu_i>0$, $i=1,\dots,\ell$.

The specific log-Student's $t$ mixtures that we will consider are
\begin{eqnarray}
&& \ell=2,\quad \nu_1=4,\quad\nu_2=12: {\rm 2LSt12} \label{model2LSt12}\\
&& \ell=2,\quad \nu_1=4,\quad\nu_2=39: {\rm 2LSt39} \label{model2LSt39}\\
&& \ell=3,\quad \nu_1=4,\quad\nu_2=12,\quad \nu_3=39: {\rm 3LSt} \label{model3LSt}\\
&& \ell=4,\quad \nu_1=4,\quad\nu_2=12,\quad \nu_3=39, \quad \nu_4=100: {\rm 4LSt} \label{model4LSt}\\
&& \ell=5,\quad \nu_1=4,\quad\nu_2=12,\quad \nu_3=39, \quad \nu_4=100,\quad \nu_5=200: {\rm 5LSt} \label{model5LSt}
\end{eqnarray}
Note that we fix the number of degrees of freedom \emph{a priori}. There are several reasons for that. First, the maximum likelihood estimation that we will perform afterwards becomes much more stable than with free degrees of freedom parameters for $\ell\geq 2$.
The choice \emph{a priori} of the parameters of the degrees of freedom in the mixtures allows us to avoid convergence problems of the numerical estimation process.
Second, with these choices, we break down the existing symmetry in the estimation parameters (concerning a scenario with free parameters of degrees of freedom in which the interchanging of the parameters of the components lead to the same distribution) so identification is achieved in this case.
And third, the versions of these distributions without logs have worked very well when studying log-growth rates of city sizes \citep{MasPueRam20}, and the log-returns of stock indices worldwide \citep{MasRam21}.

\section{The data sets}\label{data}

We use data of sales of manufacturing firms (measured in constant thousands of USA dollars, the reference year being 2005) of France (FR), Germany (DE), Italy (IT), Japan (JP), South Korea (KR) and Spain (ES), {}from the ORBIS database. For France, we have data for the years 2005-2014, for Germany, we have 2010-2014, for Italy we have 2006-2011 and for Japan, South Korea, and Spain we have 2005-2014.
ORBIS is the largest database of firms' financial data in the world. We used the 2016 edition, which still contains financial data for more than 100 million firms worldwide. It is characterized not only by the size and completeness of the data but also by the simultaneous analysis of multiple countries because the data of each country is
described in the same format. While the latest data from the 2016 edition of ORBIS is 2015, the number is limited because they are still being collected. ORBIS contains data for approximately 10 years.
In this paper, we focus on France (FR), Germany (DE), Italy (IT), Japan (JP), South Korea (KR), and
Spain (ES) as countries with sufficient data for this period.

We will use the data in three different ways in our empirical application. First, the whole data without cut-offs put us in the worst situation to obtain a good fit. Second, we will split the data into two (approximately) equally distributed subsamples: the 75\% of each of the data sets will be the in-sample data, and the other 25\% of each sample will be the out-of-sample data. We will assess the fit of the out-of-sample data into the estimated in-sample distributions. Third, for each full sample we will drop the 10\% of the bottom observations (lower tail) and the 0.1\% of the top observations (upper tail) to improve the measures of skewness and kurtosis of the log data as it is suggested in \cite{Fio20}, and more importantly, to show that the tails of the full and doubly truncated data are very difficult to model directly. We also consider all the manufacturing firms without separating by type of industry, following again the idea to put us in the worst situation to obtain a good fit, since then the sample size will be maximum and the goodness-of-fit using standard statistics will be increasingly demanding.

The descriptive statistics for the full samples can be seen in Table~\ref{descstat}.
It can be observed that the log data presents skewness and kurtosis different from that of the normal distribution, so it is guessed that a single lognormal will not be a good fit for the sales variable for these data sets.

\begin{table}[htbp]
\centering {\tiny
  \begin{tabular}{lrrrrrrrrr}
 \hline
          & \multicolumn{1}{c}{Obs} & \multicolumn{1}{c}{Mean} & \multicolumn{1}{c}{SD} & \multicolumn{1}{c}{Mean (log scale)} & \multicolumn{1}{c}{SD (log scale)} & \multicolumn{1}{c}{Skewness (log scale)} & \multicolumn{1}{c}{Kurtosis (log scale)} & \multicolumn{1}{c}{Min} & \multicolumn{1}{c}{Max} \\
    FR 2005 & 70,843 & 16,877 & 543,306 & 6.80  & 1.80  & 0.53  & 4.41  & 1     & 76,439,971 \\
    FR 2006 & 81,925 & 18,409 & 577,067 & 6.83  & 1.83  & 0.48  & 4.42  & 1     & 81,571,606 \\
    FR 2007 & 83,425 & 21,201 & 666,998 & 6.96  & 1.84  & 0.45  & 4.41  & 1     & 97,999,556 \\
    FR 2008 & 84,334 & 20,244 & 610,649 & 6.95  & 1.84  & 0.47  & 4.37  & 1     & 85,985,520 \\
    FR 2009 & 85,306 & 18,479 & 559,129 & 6.86  & 1.81  & 0.51  & 4.45  & 1     & 74,928,233 \\
    FR 2010 & 87,619 & 18,692 & 581,620 & 6.79  & 1.83  & 0.53  & 4.47  & 1     & 79,635,839 \\
    FR 2011 & 89,876 & 18,907 & 585,558 & 6.78  & 1.84  & 0.52  & 4.46  & 1     & 80,783,184 \\
    FR 2012 & 90,171 & 19,388 & 595,425 & 6.73  & 1.85  & 0.55  & 4.47  & 1     & 73,177,880 \\
    FR 2013 & 84,757 & 21,186 & 626,615 & 6.81  & 1.85  & 0.55  & 4.49  & 1     & 73,227,450 \\
    FR 2014 & 61,145 & 25,919 & 674,693 & 7.02  & 1.90  & 0.41  & 4.27  & 1     & 65,094,004 \\
    DE 2010 & 46,004 & 59,903 & 1,494,306 & 7.92  & 1.90  & 0.68  & 4.17  & 1     & 175,264,712 \\
    DE 2011 & 51,604 & 57,228 & 1,550,180 & 7.90  & 1.89  & 0.66  & 4.19  & 1     & 213,554,315 \\
    DE 2012 & 53,751 & 57,804 & 1,703,583 & 7.97  & 1.85  & 0.67  & 4.26  & 1     & 262,527,606 \\
    DE 2013 & 53,468 & 59,421 & 1,789,547 & 8.03  & 1.83  & 0.68  & 4.34  & 1     & 280,191,740 \\
    DE 2014 & 42,803 & 61,315 & 1,848,927 & 7.88  & 1.79  & 0.80  & 4.73  & 1     & 253,890,289 \\
    IT 2006 & 11,5562 & 10,467 & 142,069 & 7.37  & 1.70  & 0.04  & 4.24  & 1     & 25,595,512 \\
    IT 2007 & 126,556 & 11,684 & 170,964 & 7.24  & 1.97  & -0.35 & 4.09  & 1     & 31,901,152 \\
    IT 2008 & 129,282 & 11,165 & 162,807 & 7.16  & 1.98  & -0.33 & 4.06  & 1     & 28,226,344 \\
    IT 2009 & 129,429 & 9,475  & 146,230 & 7.01  & 1.95  & -0.31 & 4.10  & 1     & 27,295,056 \\
    IT 2010 & 129,870 & 9,725  & 147,635 & 6.98  & 1.99  & -0.30 & 4.02  & 1     & 25,831,552 \\
    IT 2011 & 129,545 & 10,231 & 151,582 & 6.99  & 2.00  & -0.29 & 3.99  & 1     & 23,953,928 \\
    JP 2005 & 17,702 & 175,035 & 2,231,910 & 8.69  & 2.32  & 0.34  & 3.18  & 2     & 179,190,023 \\
    JP 2006 & 31,174 & 119,591 & 1,883,771 & 8.39  & 2.20  & 0.31  & 3.41  & 1     & 203,553,677 \\
    JP 2007 & 38,133 & 121,330 & 2,157,447 & 8.24  & 2.21  & 0.37  & 3.50  & 1     & 262,629,781 \\
    JP 2008 & 41,802 & 106,201 & 1,793,263 & 8.26  & 2.15  & 0.35  & 3.55  & 1     & 209,271,856 \\
    JP 2009 & 46,185 & 88,161 & 1,588,579 & 8.14  & 2.10  & 0.34  & 3.69  & 1     & 203,227,599 \\
    JP 2010 & 52,892 & 95,421 & 1,837,968 & 8.12  & 2.09  & 0.38  & 3.83  & 1     & 228,481,752 \\
    JP 2011 & 60,205 & 86,197 & 1,744,006 & 8.12  & 2.04  & 0.38  & 3.89  & 1     & 226,216,104 \\
    JP 2012 & 110,434 & 48,140 & 1,209,528 & 7.67  & 1.96  & 0.45  & 3.88  & 1     & 234,351,491 \\
    JP 2013 & 150,885 & 35,090 & 1,070,805 & 7.34  & 1.85  & 0.57  & 4.28  & 1     & 249,799,825 \\
    JP 2014 & 143,327 & 32,481 & 970,026 & 7.23  & 1.87  & 0.56  & 4.27  & 1     & 226,746,495 \\
    KR 2005 & 50,320 & 19,646 & 1,062,373 & 7.43  & 1.58  & 0.25  & 4.88  & 1     & 216,488,732 \\
    KR 2006 & 59,253 & 21,474 & 700,469 & 7.49  & 1.62  & 0.21  & 5.13  & 1     & 92,315,130 \\
    KR 2007 & 57,891 & 24,301 & 807,074 & 7.61  & 1.62  & 0.22  & 4.99  & 1     & 105,232,149 \\
    KR 2008 & 59,681 & 21,772 & 730,983 & 7.47  & 1.62  & 0.22  & 4.92  & 1     & 96,303,551 \\
    KR 2009 & 55,685 & 26,378 & 884,185 & 7.61  & 1.65  & 0.20  & 4.88  & 1     & 119,359,103 \\
    KR 2010 & 58,645 & 29,856 & 922,552 & 7.75  & 1.68  & 0.22  & 4.82  & 1     & 136,262,183 \\
    KR 2011 & 61,329 & 30,909 & 944,359 & 7.77  & 1.71  & 0.19  & 4.64  & 1     & 143,255,571 \\
    KR 2012 & 64,935 & 33,416 & 1,097,013 & 7.80  & 1.72  & 0.18  & 4.64  & 1     & 187,841,967 \\
    KR 2013 & 63,589 & 35,425 & 1,188,543 & 7.85  & 1.73  & 0.17  & 4.68  & 1     & 216,688,138 \\
    KR 2014 & 52,756 & 40,135 & 1,187,518 & 7.98  & 1.76  & 0.18  & 4.55  & 1     & 187,579,357 \\
    ES 2005 & 89,211 & 5,852  & 123,123 & 6.61  & 1.64  & 0.34  & 4.17  & 1     & 21,670,942 \\
    ES 2006 & 96,021 & 7,079  & 156,785 & 6.68  & 1.69  & 0.28  & 4.18  & 1     & 27,271,390 \\
    ES 2007 & 95,436 & 8,606  & 189,066 & 6.84  & 1.70  & 0.28  & 4.15  & 1     & 31,253,101 \\
    ES 2008 & 93,670 & 8,167  & 182,975 & 6.73  & 1.73  & 0.28  & 4.20  & 1     & 35,166,726 \\
    ES 2009 & 91,181 & 7,138  & 150,541 & 6.54  & 1.75  & 0.31  & 4.20  & 1     & 26,519,791 \\
    ES 2010 & 87,691 & 7,328  & 166,231 & 6.47  & 1.77  & 0.34  & 4.20  & 1     & 29,585,314 \\
    ES 2011 & 84,282 & 7,838  & 191,230 & 6.41  & 1.82  & 0.34  & 4.13  & 1     & 35,244,632 \\
    ES 2012 & 80,444 & 8,253  & 218,209 & 6.34  & 1.86  & 0.35  & 4.05  & 1     & 38,061,289 \\
    ES 2013 & 75,871 & 9,012  & 229,237 & 6.39  & 1.89  & 0.35  & 4.01  & 1     & 38,113,298 \\
    ES 2014 & 67,527 & 8,568  & 209,032 & 6.34  & 1.88  & 0.34  & 4.03  & 1     & 31,859,934 \\
 \hline
\end{tabular}%
}
\caption{Descriptive statistics of the full data sets for sales of manufacturing firms.}
\label{descstat}
\end{table}

We show likewise the descriptive statistics of the in-sample and out-of-sample data sets in Tables~\ref{descstatis} and~\ref{descstatoos}, respectively. We can see similar descriptive statistics between these two subsamples and with regards to the full samples as well.

\begin{table}[htbp]
\centering {\tiny
    \begin{tabular}{lrrrrrrrrr}
\hline
          & \multicolumn{1}{c}{Obs} & \multicolumn{1}{c}{Mean} & \multicolumn{1}{c}{SD} & \multicolumn{1}{c}{Mean (log scale)} & \multicolumn{1}{c}{SD (log scale)} & \multicolumn{1}{c}{Skewness (log scale)} & \multicolumn{1}{c}{Kurtosis (log scale)} & \multicolumn{1}{c}{Min} & \multicolumn{1}{c}{Max} \\
    FR 2005 & 53,133 & 17,191 & 584,156 & 6.80  & 1.81  & 0.53  & 4.40  & 1     & 76,439,971 \\
    FR 2006 & 61,444 & 18,050 & 597,404 & 6.83  & 1.82  & 0.47  & 4.41  & 1     & 81,571,606 \\
    FR 2007 & 62,569 & 22,239 & 750,005 & 6.95  & 1.84  & 0.44  & 4.42  & 1     & 97,999,556 \\
    FR 2008 & 63,251 & 19,879 & 591,326 & 6.95  & 1.83  & 0.47  & 4.34  & 1     & 75,647,208 \\
    FR 2009 & 63,980 & 18,438 & 522,725 & 6.86  & 1.81  & 0.51  & 4.48  & 1     & 69,750,990 \\
    FR 2010 & 65,715 & 18,076 & 603,392 & 6.78  & 1.82  & 0.51  & 4.47  & 1     & 79,635,839 \\
    FR 2011 & 67,407 & 18,762 & 578,352 & 6.78  & 1.85  & 0.52  & 4.48  & 1     & 80,783,184 \\
    FR 2012 & 67,629 & 19,879 & 601,232 & 6.73  & 1.85  & 0.55  & 4.48  & 1     & 71,518,074 \\
    FR 2013 & 63,568 & 21,665 & 656,044 & 6.81  & 1.85  & 0.56  & 4.48  & 1     & 73,227,450 \\
    FR 2014 & 45,859 & 28,518 & 749,215 & 7.04  & 1.91  & 0.42  & 4.31  & 1     & 65,094,004 \\
    DE 2010 & 34,503 & 61,821 & 1,595,579 & 7.92  & 1.90  & 0.67  & 4.21  & 1     & 175,264,712 \\
    DE 2011 & 38,703 & 59,226 & 1,612,308 & 7.90  & 1.90  & 0.66  & 4.20  & 1     & 213,554,315 \\
    DE 2012 & 40,314 & 59,983 & 1,849,078 & 7.97  & 1.86  & 0.67  & 4.26  & 1     & 262,527,606 \\
    DE 2013 & 40,101 & 66,053 & 2,027,592 & 8.03  & 1.83  & 0.69  & 4.40  & 1     & 280,191,740 \\
    DE 2014 & 32,103 & 53,470 & 1,347,504 & 7.88  & 1.79  & 0.80  & 4.66  & 1     & 159,736,788 \\
    IT 2006 & 86,672 & 10,574 & 133,842 & 7.37  & 1.70  & 0.03  & 4.27  & 1     & 15,364,152 \\
    IT 2007 & 94,917 & 11,688 & 178,721 & 7.24  & 1.97  & -0.35 & 4.09  & 1     & 31,901,152 \\
    IT 2008 & 96,962 & 11,338 & 163,229 & 7.16  & 1.98  & -0.33 & 4.05  & 1     & 28,226,344 \\
    IT 2009 & 97,072 & 9,646  & 161,343 & 7.01  & 1.95  & -0.31 & 4.10  & 1     & 27,295,056 \\
    IT 2010 & 97,403 & 9,930  & 160,228 & 6.98  & 1.98  & -0.29 & 4.03  & 1     & 25,831,552 \\
    IT 2011 & 97,159 & 10,033 & 152,013 & 6.99  & 2.01  & -0.30 & 3.98  & 1     & 23,953,928 \\
    JP 2005 & 13,277 & 175,571 & 2,351,840 & 8.68  & 2.32  & 0.34  & 3.17  & 2     & 179,190,023 \\
    JP 2006 & 23,381 & 126,250 & 2,047,376 & 8.39  & 2.21  & 0.32  & 3.44  & 1     & 203,553,677 \\
    JP 2007 & 28,600 & 110,996 & 1,474,582 & 8.25  & 2.21  & 0.38  & 3.49  & 1     & 119,908,437 \\
    JP 2008 & 31,352 & 103,557 & 1,894,180 & 8.25  & 2.15  & 0.34  & 3.52  & 1     & 209,271,856 \\
    JP 2009 & 34,639 & 83,383 & 1,539,591 & 8.15  & 2.10  & 0.33  & 3.67  & 1     & 203,227,599 \\
    JP 2010 & 39,669 & 87,976 & 1,730,390 & 8.12  & 2.08  & 0.38  & 3.79  & 1     & 228,481,752 \\
    JP 2011 & 45,154 & 86,109 & 1,772,454 & 8.11  & 2.04  & 0.38  & 3.87  & 1     & 226,216,104 \\
    JP 2012 & 82,826 & 49,719 & 1,321,483 & 7.67  & 1.96  & 0.44  & 3.88  & 1     & 234,351,491 \\
    JP 2013 & 113,164 & 35,004 & 1,085,891 & 7.34  & 1.85  & 0.57  & 4.29  & 1     & 249,799,825 \\
    JP 2014 & 107,496 & 34,279 & 845,356 & 7.23  & 1.88  & 0.58  & 4.32  & 1     & 110,965,775 \\
    KR 2005 & 37,740 & 19,940 & 1,161,412 & 7.43  & 1.59  & 0.26  & 4.91  & 1     & 216,488,732 \\
    KR 2006 & 44,440 & 21,059 & 727,199 & 7.48  & 1.62  & 0.21  & 5.07  & 1     & 92,315,130 \\
    KR 2007 & 43,419 & 25,294 & 831,783 & 7.62  & 1.62  & 0.24  & 5.01  & 1     & 105,232,149 \\
    KR 2008 & 44,761 & 22,942 & 777,121 & 7.48  & 1.62  & 0.23  & 4.91  & 1     & 96,303,551 \\
    KR 2009 & 41,764 & 26,912 & 826,475 & 7.61  & 1.65  & 0.22  & 4.90  & 1     & 86,447,464 \\
    KR 2010 & 43,984 & 31,489 & 945,086 & 7.76  & 1.68  & 0.24  & 4.83  & 1     & 136,262,183 \\
    KR 2011 & 45,997 & 28,815 & 894,458 & 7.77  & 1.71  & 0.17  & 4.61  & 1     & 143,255,571 \\
    KR 2012 & 48,702 & 37,139 & 1,246,562 & 7.80  & 1.73  & 0.17  & 4.67  & 1     & 187,841,967 \\
    KR 2013 & 47,692 & 32,814 & 875,328 & 7.85  & 1.74  & 0.17  & 4.64  & 1     & 106,892,322 \\
    KR 2014 & 39,567 & 44,157 & 1,343,318 & 7.98  & 1.76  & 0.20  & 4.61  & 1     & 187,579,357 \\
    ES 2005 & 66,909 & 5,813  & 115,435 & 6.61  & 1.63  & 0.36  & 4.20  & 1     & 21,670,942 \\
    ES 2006 & 72,016 & 7,205  & 172,458 & 6.67  & 1.69  & 0.29  & 4.16  & 1     & 27,271,390 \\
    ES 2007 & 71,577 & 8,618  & 197,828 & 6.84  & 1.70  & 0.28  & 4.14  & 1     & 31,253,101 \\
    ES 2008 & 70,253 & 8,038  & 172,697 & 6.74  & 1.73  & 0.27  & 4.19  & 1     & 35,166,726 \\
    ES 2009 & 68,386 & 6,959  & 148,637 & 6.54  & 1.74  & 0.30  & 4.19  & 1     & 26,519,791 \\
    ES 2010 & 65,769 & 7,315  & 169,762 & 6.47  & 1.77  & 0.33  & 4.20  & 1     & 29,585,314 \\
    ES 2011 & 63,212 & 8,587  & 218,795 & 6.41  & 1.82  & 0.35  & 4.19  & 1     & 35,244,632 \\
    ES 2012 & 60,333 & 7,949  & 189,833 & 6.34  & 1.86  & 0.35  & 4.05  & 1     & 33,693,747 \\
    ES 2013 & 56,904 & 9,613  & 258,885 & 6.39  & 1.90  & 0.36  & 4.03  & 1     & 38,113,298 \\
    ES 2014 & 50,646 & 8,778  & 219,055 & 6.34  & 1.88  & 0.35  & 4.03  & 1     & 31,859,934 \\
\hline
\end{tabular}%
}
\caption{Descriptive statistics of the in-sample data sets for sales of manufacturing firms.}
\label{descstatis}
\end{table}

\begin{table}[htbp]
\centering {\tiny
  \begin{tabular}{lrrrrrrrrr}
  \hline
          & \multicolumn{1}{c}{Obs} & \multicolumn{1}{c}{Mean} & \multicolumn{1}{c}{SD} & \multicolumn{1}{c}{Mean (log scale)} & \multicolumn{1}{c}{SD (log scale)} & \multicolumn{1}{c}{Skewness (log scale)} & \multicolumn{1}{c}{Kurtosis (log scale)} & \multicolumn{1}{c}{Min} & \multicolumn{1}{c}{Max} \\
    FR 2005 & 17,710 & 15,935 & 396,245 & 6.79  & 1.79  & 0.53  & 4.43  & 1     & 41,441,833 \\
    FR 2006 & 20,481 & 19,485 & 511,237 & 6.84  & 1.84  & 0.50  & 4.46  & 1     & 54,793,890 \\
    FR 2007 & 20,856 & 18,088 & 303,342 & 6.97  & 1.85  & 0.49  & 4.38  & 1     & 18,807,541 \\
    FR 2008 & 21,083 & 21,338 & 665,274 & 6.94  & 1.85  & 0.49  & 4.45  & 1     & 85,985,520 \\
    FR 2009 & 21,326 & 18,604 & 656,353 & 6.87  & 1.81  & 0.50  & 4.36  & 1     & 74,928,233 \\
    FR 2010 & 21,904 & 20,541 & 510,770 & 6.81  & 1.84  & 0.58  & 4.48  & 1     & 44,427,809 \\
    FR 2011 & 22,469 & 19,342 & 606,678 & 6.77  & 1.82  & 0.50  & 4.39  & 1     & 54,493,893 \\
    FR 2012 & 22,542 & 17,914 & 577,661 & 6.74  & 1.85  & 0.53  & 4.43  & 1     & 73,177,880 \\
    FR 2013 & 21,189 & 19,748 & 528,601 & 6.81  & 1.84  & 0.53  & 4.50  & 1     & 48,443,644 \\
    FR 2014 & 15,286 & 18,119 & 369,867 & 6.99  & 1.87  & 0.38  & 4.16  & 1     & 30,393,794 \\
    DE 2010 & 11,501 & 54,149 & 1,137,668 & 7.91  & 1.91  & 0.70  & 4.05  & 1     & 87,328,880 \\
    DE 2011 & 12,901 & 51,236 & 1,346,747 & 7.89  & 1.88  & 0.67  & 4.15  & 1     & 139,490,185 \\
    DE 2012 & 13,437 & 51,269 & 1,162,531 & 7.96  & 1.85  & 0.66  & 4.24  & 1     & 96,303,003 \\
    DE 2013 & 13,367 & 39,527 & 690,000 & 8.03  & 1.82  & 0.65  & 4.15  & 1     & 71,738,898 \\
    DE 2014 & 10,700 & 84,852 & 2,868,302 & 7.87  & 1.79  & 0.77  & 4.93  & 1     & 253,890,289 \\
    IT 2006 & 28,890 & 10,145 & 164,300 & 7.37  & 1.69  & 0.08  & 4.15  & 1     & 25,595,512 \\
    IT 2007 & 31,639 & 11,671 & 145,231 & 7.24  & 1.98  & -0.36 & 4.09  & 1     & 15,024,101 \\
    IT 2008 & 32,320 & 10,646 & 161,532 & 7.16  & 1.97  & -0.33 & 4.07  & 1     & 21,903,956 \\
    IT 2009 & 32,357 & 8,963  & 86,241 & 7.01  & 1.96  & -0.30 & 4.11  & 1     & 6,371,920 \\
    IT 2010 & 32,467 & 9,112  & 100,819 & 6.98  & 2.00  & -0.33 & 3.99  & 1     & 7,751,384 \\
    IT 2011 & 32,386 & 10,824 & 150,284 & 7.00  & 1.99  & -0.25 & 4.05  & 1     & 15,998,212 \\
    JP 2005 & 4,425  & 173,424 & 1,825,613 & 8.71  & 2.33  & 0.33  & 3.20  & 4     & 75,760,893 \\
    JP 2006 & 7,793  & 99,613 & 1,272,289 & 8.38  & 2.20  & 0.31  & 3.31  & 1     & 88,980,728 \\
    JP 2007 & 9,533  & 152,334 & 3,477,804 & 8.21  & 2.21  & 0.36  & 3.54  & 1     & 262,629,781 \\
    JP 2008 & 10,450 & 114,136 & 1,448,967 & 8.28  & 2.17  & 0.39  & 3.63  & 1     & 79,159,089 \\
    JP 2009 & 11,546 & 102,495 & 1,727,221 & 8.11  & 2.10  & 0.38  & 3.76  & 1     & 96,177,440 \\
    JP 2010 & 13,223 & 117,755 & 2,128,241 & 8.13  & 2.10  & 0.37  & 3.95  & 1     & 112,063,118 \\
    JP 2011 & 15,051 & 86,463 & 1,655,787 & 8.12  & 2.03  & 0.37  & 3.94  & 1     & 117,661,387 \\
    JP 2012 & 27,608 & 43,403 & 782,856 & 7.69  & 1.96  & 0.47  & 3.88  & 1     & 96,028,373 \\
    JP 2013 & 37,721 & 35,346 & 1,024,227 & 7.33  & 1.85  & 0.57  & 4.23  & 1     & 115,142,941 \\
    JP 2014 & 35,831 & 27,087 & 1,272,770 & 7.21  & 1.84  & 0.49  & 4.11  & 1     & 226,746,495 \\
    KR 2005 & 12,580 & 18,764 & 684,080 & 7.44  & 1.58  & 0.23  & 4.82  & 1     & 61,334,641 \\
    KR 2006 & 14,813 & 22,720 & 613,350 & 7.49  & 1.62  & 0.20  & 5.30  & 1     & 61,170,593 \\
    KR 2007 & 14,472 & 21,323 & 727,944 & 7.61  & 1.61  & 0.14  & 4.94  & 1     & 81,495,922 \\
    KR 2008 & 14,920 & 18,262 & 570,614 & 7.44  & 1.62  & 0.18  & 4.95  & 1     & 63,307,942 \\
    KR 2009 & 13,921 & 24,777 & 1,038,280 & 7.60  & 1.66  & 0.15  & 4.81  & 1     & 119,359,103 \\
    KR 2010 & 14,661 & 24,958 & 851,390 & 7.74  & 1.68  & 0.15  & 4.81  & 1     & 99,215,436 \\
    KR 2011 & 15,332 & 37,191 & 1,080,330 & 7.79  & 1.71  & 0.25  & 4.73  & 1     & 99,871,779 \\
    KR 2012 & 16,233 & 22,245 & 389,608 & 7.78  & 1.72  & 0.18  & 4.56  & 1     & 33,300,928 \\
    KR 2013 & 15,897 & 43,257 & 1,830,865 & 7.86  & 1.72  & 0.17  & 4.81  & 1     & 216,688,138 \\
    KR 2014 & 13,189 & 28,070 & 476,598 & 7.96  & 1.77  & 0.15  & 4.37  & 1     & 37,704,286 \\
    ES 2005 & 22,302 & 5,969  & 143,746 & 6.62  & 1.64  & 0.31  & 4.08  & 1     & 18,724,815 \\
    ES 2006 & 24,005 & 6,700  & 95,404 & 6.69  & 1.69  & 0.26  & 4.24  & 1     & 7,462,464 \\
    ES 2007 & 23,859 & 8,570  & 159,931 & 6.85  & 1.71  & 0.28  & 4.20  & 1     & 20,185,108 \\
    ES 2008 & 23,417 & 8,553  & 210,826 & 6.71  & 1.73  & 0.30  & 4.24  & 1     & 26,376,760 \\
    ES 2009 & 22,795 & 7,673  & 156,115 & 6.54  & 1.75  & 0.32  & 4.25  & 1     & 18,450,121 \\
    ES 2010 & 21,922 & 7,367  & 155,158 & 6.47  & 1.76  & 0.37  & 4.20  & 1     & 17,159,719 \\
    ES 2011 & 21,070 & 5,591  & 51,525 & 6.42  & 1.80  & 0.32  & 3.95  & 1     & 3,000,598 \\
    ES 2012 & 20,111 & 9,162  & 286,974 & 6.34  & 1.87  & 0.37  & 4.06  & 1     & 38,061,289 \\
    ES 2013 & 18,967 & 7,211  & 95,547 & 6.38  & 1.88  & 0.34  & 3.93  & 1     & 6,976,931 \\
    ES 2014 & 16,881 & 7,936  & 175,566 & 6.33  & 1.88  & 0.33  & 4.02  & 1     & 20,480,750 \\
\hline
\end{tabular}%
}
\caption{Descriptive statistics of the out-of-sample data sets for sales of manufacturing firms.}
\label{descstatoos}
\end{table}

And finally, we show the descriptive statistics of the doubly truncated (tt) data sets in the way explained before, in Table~\ref{descstattt}. Now the descriptive statistics do vary somehow, in particular, the skewness and kurtosis of the log data are slightly closer to the ones of the normal distribution, as suggested in \cite{Fio20}.

\begin{table}[htbp]
\centering {\tiny
   \begin{tabular}{lrrrrrrrrr}
\hline
          & \multicolumn{1}{c}{Obs} & \multicolumn{1}{c}{Mean} & \multicolumn{1}{c}{SD} & \multicolumn{1}{c}{Mean (log scale)} & \multicolumn{1}{c}{SD (log scale)} & \multicolumn{1}{c}{Skewness (log scale)} & \multicolumn{1}{c}{Kurtosis (log scale)} & \multicolumn{1}{c}{Min} & \multicolumn{1}{c}{Max} \\
    FR 2005 & 63,689 & 8,745  & 53,959 & 7.10  & 1.56  & 1.04  & 4.18  & 120   & 1,653,698 \\
    FR 2006 & 73,651 & 9,311  & 57,868 & 7.15  & 1.57  & 1.02  & 4.16  & 122   & 1,776,995 \\
    FR 2007 & 74,999 & 10,752 & 66,746 & 7.28  & 1.58  & 1.01  & 4.11  & 136   & 2,122,012 \\
    FR 2008 & 75,817 & 10,529 & 65,992 & 7.26  & 1.58  & 1.01  & 4.09  & 135   & 2,148,030 \\
    FR 2009 & 76,690 & 9,475  & 59,544 & 7.17  & 1.56  & 1.04  & 4.19  & 128   & 1,929,253 \\
    FR 2010 & 78,770 & 9,336  & 60,524 & 7.10  & 1.58  & 1.06  & 4.24  & 118   & 1,984,206 \\
    FR 2011 & 80,799 & 9,522  & 62,752 & 7.10  & 1.59  & 1.06  & 4.23  & 115   & 2,001,534 \\
    FR 2012 & 81,064 & 9,429  & 62,933 & 7.05  & 1.60  & 1.07  & 4.26  & 110   & 2,023,960 \\
    FR 2013 & 76,197 & 10,343 & 70,060 & 7.12  & 1.60  & 1.07  & 4.27  & 118   & 2,290,181 \\
    FR 2014 & 54,969 & 12,783 & 87,566 & 7.36  & 1.62  & 0.94  & 4.07  & 125   & 3,214,342 \\
    DE 2010 & 41,358 & 31,525 & 177,373 & 8.22  & 1.69  & 1.06  & 3.83  & 347   & 6,079,053 \\
    DE 2011 & 46,393 & 30,563 & 170,770 & 8.20  & 1.68  & 1.06  & 3.88  & 336   & 5,876,894 \\
    DE 2012 & 48,323 & 30,492 & 169,305 & 8.27  & 1.64  & 1.08  & 3.95  & 389   & 5,358,436 \\
    DE 2013 & 48,069 & 31,616 & 175,947 & 8.33  & 1.62  & 1.11  & 4.08  & 414   & 5,348,941 \\
    DE 2014 & 38,481 & 29,379 & 190,361 & 8.17  & 1.59  & 1.21  & 4.55  & 370   & 6,798,963 \\
    IT 2006 & 103,891 & 8,412  & 34,530 & 7.69  & 1.41  & 0.72  & 3.46  & 216   & 948,514 \\
    IT 2007 & 113,775 & 9,224  & 38,979 & 7.65  & 1.52  & 0.53  & 3.22  & 123   & 1,061,414 \\
    IT 2008 & 116,225 & 8,757  & 36,921 & 7.58  & 1.53  & 0.54  & 3.25  & 113   & 991,101 \\
    IT 2009 & 116,358 & 7,437  & 32,324 & 7.42  & 1.51  & 0.58  & 3.32  & 101   & 954,812 \\
    IT 2010 & 116,754 & 7,601  & 33,259 & 7.40  & 1.54  & 0.56  & 3.29  & 94    & 932,925 \\
    IT 2011 & 116,462 & 7,948  & 35,135 & 7.41  & 1.56  & 0.57  & 3.28  & 92    & 977,324 \\
    JP 2005 & 15,915 & 135,714 & 895,942 & 9.09  & 2.03  & 0.64  & 3.22  & 336   & 23,782,870 \\
    JP 2006 & 28,026 & 84,891 & 613,838 & 8.78  & 1.91  & 0.67  & 3.43  & 278   & 18,723,918 \\
    JP 2007 & 34,282 & 81,020 & 620,897 & 8.63  & 1.92  & 0.73  & 3.52  & 252   & 18,946,444 \\
    JP 2008 & 37,581 & 71,503 & 541,412 & 8.64  & 1.86  & 0.74  & 3.57  & 277   & 16,629,918 \\
    JP 2009 & 41,521 & 58,489 & 441,659 & 8.52  & 1.80  & 0.78  & 3.73  & 265   & 13,789,223 \\
    JP 2010 & 47,551 & 60,095 & 473,146 & 8.50  & 1.79  & 0.84  & 3.86  & 273   & 15,240,274 \\
    JP 2011 & 54,125 & 52,758 & 407,409 & 8.48  & 1.75  & 0.84  & 3.88  & 287   & 13,503,043 \\
    JP 2012 & 99,281 & 27,926 & 200,081 & 8.02  & 1.69  & 0.87  & 3.83  & 212   & 6,649,198 \\
    JP 2013 & 135,647 & 18,290 & 138,069 & 7.66  & 1.61  & 1.01  & 4.24  & 190   & 4,850,462 \\
    JP 2014 & 128,852 & 17,143 & 132,121 & 7.55  & 1.62  & 1.01  & 4.24  & 165   & 4,539,722 \\
    KR 2005 & 45,238 & 8,876  & 47,262 & 7.71  & 1.33  & 0.91  & 4.17  & 274   & 1,661,595 \\
    KR 2006 & 53,269 & 10,329 & 60,393 & 7.78  & 1.35  & 0.94  & 4.30  & 284   & 2,234,504 \\
    KR 2007 & 52,045 & 11,402 & 64,613 & 7.91  & 1.34  & 0.93  & 4.28  & 321   & 2,361,544 \\
    KR 2008 & 53,654 & 9,990  & 57,104 & 7.77  & 1.34  & 0.92  & 4.26  & 275   & 2,057,356 \\
    KR 2009 & 50,062 & 12,232 & 72,722 & 7.91  & 1.37  & 0.91  & 4.27  & 297   & 2,402,065 \\
    KR 2010 & 52,723 & 15,007 & 88,601 & 8.06  & 1.40  & 0.93  & 4.22  & 335   & 3,082,866 \\
    KR 2011 & 55,135 & 15,395 & 86,329 & 8.09  & 1.42  & 0.88  & 4.04  & 330   & 3,171,590 \\
    KR 2012 & 58,378 & 16,067 & 88,702 & 8.12  & 1.42  & 0.89  & 4.05  & 339   & 3,006,220 \\
    KR 2013 & 57,168 & 17,291 & 95,372 & 8.17  & 1.43  & 0.90  & 4.08  & 351   & 3,126,179 \\
    KR 2014 & 47,429 & 20,358 & 111,557 & 8.30  & 1.46  & 0.87  & 3.92  & 387   & 3,765,714 \\
    ES 2005 & 80,201 & 4,229  & 19,509 & 6.90  & 1.39  & 0.89  & 3.91  & 113   & 537,267 \\
    ES 2006 & 86,323 & 4,923  & 24,157 & 6.98  & 1.43  & 0.88  & 3.89  & 112   & 719,247 \\
    ES 2007 & 85,797 & 5,934  & 29,349 & 7.15  & 1.44  & 0.87  & 3.86  & 129   & 874,898 \\
    ES 2008 & 84,210 & 5,637  & 28,719 & 7.04  & 1.46  & 0.89  & 3.90  & 114   & 856,863 \\
    ES 2009 & 81,972 & 4,907  & 25,216 & 6.85  & 1.47  & 0.92  & 3.95  & 93    & 734,985 \\
    ES 2010 & 78,835 & 4,945  & 26,261 & 6.79  & 1.50  & 0.94  & 3.99  & 86    & 824,988 \\
    ES 2011 & 75,770 & 5,156  & 27,937 & 6.74  & 1.54  & 0.93  & 3.94  & 76    & 867,938 \\
    ES 2012 & 72,320 & 5,324  & 30,004 & 6.68  & 1.58  & 0.92  & 3.89  & 67    & 880,194 \\
    ES 2013 & 68,209 & 5,865  & 33,454 & 6.73  & 1.61  & 0.92  & 3.85  & 67    & 991,864 \\
    ES 2014 & 60,707 & 5,514  & 31,740 & 6.68  & 1.60  & 0.92  & 3.88  & 65    & 927,473 \\
\hline
\end{tabular}%
}
\caption{Descriptive statistics of the doubly truncated data sets for sales of manufacturing firms.}
\label{descstattt}
\end{table}

\section{Empirical results}\label{results}

We will summarize in this Section the results of fitting different models to each different type of data, separating them into different Subsections. But there are aspects of the method followed that are common to all the analyses. First, we have only taken into account distributions that could have the regularity conditions of, for example, \cite{Kie78,BasMcL85,NewMcF94,CasBer02,AtiGarMunVil07} and avoiding distributions whose probability density is not three times continuously differentiable in the parameters and distributions that have not a CDF expressible in terms of elementary or special functions known to date. Also, the maximum likelihood estimation (MLE) that we have carried out has been done using the command {\tt mle} of {\sc MATLAB}$^\circledR$, which relies on the Nelder--Mead simplex algorithm \citep{NelMea65}. We have chosen this to treat all the distributions on an equal footing, instead of choosing an EM algorithm for the mixture models \citep{McLPee00} and using the above-cited command {\tt mle} for the other distributions. Independently of the estimations obtained by the {\tt mle} command, we have checked that the estimations correspond to a maximum of the log-likelihood by varying the initial starting points.
We have checked as well that the maxima are so around intervals centred on the estimate and with a width of eight standard errors.
The standard errors are computed following the procedure of \cite{EfrHin78} and \cite{McCVin03}.

To assess the goodness-of-fit of the different models to describe the data, we have computed standard Kolmogorov--Smirnov (KS), Cr\'amer--von Mises (CM), and Anderson--Darling (AD) statistics.
Let us recall how these statistics are computed.
\begin{itemize}
\item[$\bullet$] The \emph{Kolmogorov-Smirnov (KS) statistic} \cite[]{KolmogorovAN1933} compares the empirical distribution function
${\rm cdf}_n(x)=\frac{1}{n}\sum_{i=1}^n\mathds{1}_{(0,x]}(x_i)$
with the distribution function of the fitted model ${\rm cdf}_{B}(x;\hat{\theta})$ and is given by the maximal deviance
\begin{equation*}
KS=\sup_{x\in (0,\infty)}|{\rm cdf}_n(x)-{\rm cdf}_B(x;\hat{\theta})|
\end{equation*}
and $n$ is the sample size.
\item[$\bullet$] The \emph{Cram\'er--von Mises (CM) statistic} \cite[]{Cra28,Mis28} is defined by
\begin{align*}
CM&=n\int_{0}^{\infty}({\rm cdf}_n(x)-{\rm cdf}_{B}(x;\hat{\theta}))^2\mathrm{d}{\rm cdf}_{B}(x;\hat{\theta})\\
&=\frac{1}{12n}
+\sum_{i=1}^n\left(\frac{2i-1}{2n}-{\rm cdf}_{B}(x_{(i)};\hat{\theta})\right)^2,
\end{align*}
\item[$\bullet$] The \emph{Anderson-Darling (AD) statistic} \cite[]{AndDar54} is defined by
\begin{align*}
AD&=n\int_{0}^{\infty}\frac{({\rm cdf}_n(x)-{\rm cdf}_{B}(x;\hat{\theta}))^2}{{\rm cdf}_{B}(x;\hat{\theta})(1-{\rm cdf}_{B}(x;\hat{\theta}))}\mathrm{d}{\rm cdf}_{B}(x;\hat{\theta})\\
&=-n-\sum_{i=1}^n\frac{2i-1}{n}\left(\log {\rm cdf}_{B}(x_{(i)};\hat{\theta})+\log\left(1-{\rm cdf}_{B}(x_{(n+1-i)};\hat{\theta})\right)\right),
\end{align*}
where $x_{(1)}\leq\ldots\leq x_{(n)}$ are the observed ordered data.
\end{itemize}
The smaller the value of these statistics or distances, the most favoured the evaluated model.

In order to select models, we resort to standard information criteria very well adapted to the ML estimation. They are:
\begin{itemize}
\item[$\bullet$] The \emph{Akaike Information Criterion (AIC)} \citep{Aka74,BurAnd02,BurAnd04}, defined as
    $$
    AIC=2k-2\ln L^*
    $$
    where $k$ is the number of parameters of the distribution and $\ln L^*$ is the corresponding (maximum) log-likelihood.
    The minimum value of AIC corresponds (asymptotically) to the minimum value of the Kullback--Leibler divergence, so a model with the lowest AIC is selected from among the competitors.
\item[$\bullet$] The \emph{Bayesian or Schwarz Information Criterion (BIC)} \citep{Sch78,BurAnd02,BurAnd04}, defined as
    $$
    BIC=k \ln(n)-2 \ln L^*
    $$
    where $k$ is the number of parameters of the distribution, $n$ the sample size, and $\ln L^*$ is as before. The BIC penalizes more heavily the number of parameters used than does the AIC.
    The model with the lowest BIC is selected according to this criterion.
\item[$\bullet$] The \emph{Hannan--Quinn Information Criterion (HQC)} \citep{HanQui79,BurAnd02,BurAnd04}, defined as
    $$HQC=2 k \ln(\ln(n))-2 \ln L^*
    $$
    where $k$ is the number of parameters of the distribution, $n$ the sample size, and $\ln L^*$ is as before. The HQC implements an intermediate penalization of the number of parameters when compared to the AIC and BIC. The model with the lowest HQC is selected according to this criterion.
\end{itemize}

\subsection{Full samples' results}\label{fullsamplesresults}

We start showing the empirical results for the full samples. The used distributions can be estimated almost always, with the following exceptions: The GB2 and LNSNP for JP 2005, the 5LN, 3LL, and 4LL for IT 2007-2011, the 4LL for JP 2014, the 5LL for IT 2007-2010 and JP 2014, the 3LSt for JP 2010, and the 5LSt for IT 2007-2011. For the rest of the cases, all estimations correspond to a well-behaved maximum with corresponding standard errors that are shown on different sheets of a supplementary Excel book.\footnote{This file is available at \href{https://doi.org/10.7910/DVN/PHHJM3}
{https://doi.org/10.7910/DVN/PHHJM3}.}

Next, the computation of the goodness-of-fit statistics KS, CM, and AD yield the summarized outcomes in Table~\ref{KSCMADfullsample}.

\begin{table}[htbp]
\centering {
    \begin{tabular}{lrrrrrrr}
    \hline
           & \multicolumn{1}{l}{Dist. Eq.} & \multicolumn{1}{l}{$k$} & \multicolumn{1}{l}{Min KS} &       & \multicolumn{1}{l}{Min CM} &       & \multicolumn{1}{l}{Min AD} \\
    LN    & (\ref{modelLN})& 2 & 0     &       & 0     &       & 0 \\
    DPLN  & (\ref{modelDPLN})&4 &  0     &       & 0     &       & 0 \\
    GB2   & (\ref{modelGB2})&4 & 0     &       & 0     &       & 0 \\
    LNSNP & (\ref{modelLNSNP})& 6 & 0     &       & 0     &       & 0 \\
    2LN   & (\ref{model2LN}) & 5 & 0     &       & 0     &       & 0 \\
    3LN   & (\ref{model3LN})& 8 & 5     &       & 2     &       & 4 \\
    4LN   & (\ref{model4LN})& 11 & 2     &       & 3     &       & 3 \\
    5LN   & (\ref{model5LN}) &14 & 5     &       & 9     &       & 9 \\
    2LL   & (\ref{model2LL})& 5 &  0     &       & 0     &       & 0 \\
    3LL   & (\ref{model3LL})& 8 &  1     &       & 0     &       & 0 \\
    4LL   & (\ref{model4LL})& 11 &11    &       & 3     &       & 1 \\
    5LL   & (\ref{model5LL})& 14 & 17    &       & 19    &       & 16 \\
    2LSt12 & (\ref{model2LSt12})& 5& 0     &       & 0     &       & 0 \\
    2LSt39 & (\ref{model2LSt39})& 5 & 0     &       & 0     &       & 0 \\
    3LSt  & (\ref{model3LSt})& 8 &  1     &       & 2     &       & 0 \\
    4LSt  & (\ref{model4LSt})& 11 & 4     &       & 4     &       & 5 \\
    5LSt  & (\ref{model5LSt})& 14 & 5     &       & 9     &       & 13 \\
    Total & & & 51    &       & 51    &       & 51 \\
 \hline
\end{tabular}%
}
\caption{Summary of the results of the minimum KS, CM, AD tests' statistics for each model for the full samples' data sets.}
\label{KSCMADfullsample}
\end{table}

{}From Table~\ref{KSCMADfullsample} we can see that several distributions never correspond to the minimum of either KS, CM, or AD, amongst them, the lognormal distribution, and several others. We will consider the 51 possibilities times three statistics each, which amounts to 153 cases in total, and also in what follows. The top-six of minimum statistics models are the 5LL (52 times out of 153), 5LSt (27), 5LN (23), 4LL (15), 4LSt (13), 3LN (11), showing that the mixtures yield the best results and with an increasing number of components in the mixture. This is reasonable, as with more components the fit is supposed to be better.
This is our next task to see how these models are selected according to information criteria, that penalize the number of parameters.

We show in Table~\ref{AICBICHQCfullsamples} the summarized results for the full samples' case. We can see that the five-top selected models are the 4LSt (45 times out of 153), the 5LN (31), the 3LN (23), the 4LN (18), and the 5LSt (13).
The non-mixture models are never selected, but some mixtures like the 2LL, 3LL, 2LSt12, 2LSt39, not either.
One could argue that there is a slight tendency to have better and better information criteria with a higher number of components. We conjecture that this is due to the difficulty of modelling the tails of the distribution of the full data sets, as it happened for example in \cite{Ram22} and references therein when studying the upper tail of world billionaires' data. We will consider the doubly truncated data sets in the way explained before in Subsection~\ref{ttsamplesresults} and then we will see the differences in the selected models concerning those in this Subsection.

\begin{table}[htbp]
\centering {
    \begin{tabular}{lrrrrrrr}
    \hline
          & \multicolumn{1}{l}{Dist. Eq.} & \multicolumn{1}{l}{$k$} & \multicolumn{1}{l}{Sel AIC} &       & \multicolumn{1}{l}{Sel BIC} &       & \multicolumn{1}{l}{Sel HQC} \\
    LN    & (\ref{modelLN})&2 & 0     &       & 0     &       & 0 \\
    DPLN  & (\ref{modelDPLN})&4 & 0     &       & 0     &       & 0 \\
    GB2   & (\ref{modelGB2})&4 & 0     &       & 0     &       & 0 \\
    LNSNP & (\ref{modelLNSNP})&6 & 0     &       & 0     &       & 0 \\
    2LN   & (\ref{model2LN})&5 & 1     &       & 3     &       & 3 \\
    3LN   & (\ref{model3LN})&8 & 2     &       & 17    &       & 4 \\
    4LN   & (\ref{model4LN})& 11 & 5     &       & 5     &       & 8 \\
    5LN   & (\ref{model5LN})& 14 & 19    &       & 1     &       & 11 \\
    2LL   & (\ref{model2LL})& 5 & 0     &       & 0     &       & 0 \\
    3LL   & (\ref{model3LL})& 8 & 0     &       & 0     &       & 0 \\
    4LL   & (\ref{model4LL})& 11 & 3     &       & 3     &       & 3 \\
    5LL   & (\ref{model5LL})& 14 &  2     &       & 1     &       & 1 \\
    2LSt12 & (\ref{model2LSt12})& 5 & 0     &       & 0     &       & 0 \\
    2LSt39 & (\ref{model2LSt39})& 5 & 0     &       & 0     &       & 0 \\
    3LSt  & (\ref{model3LSt})& 8 & 0     &       & 3     &       & 0 \\
    4LSt  & (\ref{model4LSt})& 11 & 11    &       & 18    &       & 16 \\
    5LSt  & (\ref{model5LSt})& 14 & 8     &       & 0     &       & 5 \\
    Total & & & 51    &       & 51    &       & 51 \\
 \hline
\end{tabular}%
}
\caption{Summary of the results of the selection of the distributions for the full samples by information criteria AIC, BIC, HQC.}
\label{AICBICHQCfullsamples}
\end{table}



\subsection{A study of in-sample and out-of-sample assessment}\label{insampleoutofsample}

In this subsection, we study the descriptive power of different models when we separate each full sample into two subsamples: in-sample (75\% of the full data) and out-of-sample (25\% of the full data), the separation being made in an identically distributed way. We have checked whether the two sub-samples so obtained come from the same distribution with standard KS, CM, and AD tests, and the global result is that at the 5\% level the null hypothesis of coming {}from the same distribution is not rejected 143 out of 153 times, so it becomes a very reasonable assumption. The details are shown on a sheet of the cited supplementary Excel book.\footnote{Again, this file is available at \href{https://doi.org/10.7910/DVN/PHHJM3}
{https://doi.org/10.7910/DVN/PHHJM3}.}

Then, we have fitted the 17 distributions used in this paper to the in-sample data sets and computed the standard errors as before. The cases that we have not been able to estimate are the following: The GB2 and LNSNP for JP 2005, the 4LN for IT 2007-2010, the 5LN for IT 2007-2011, the 3LL for IT 2007-2011, the 4LL for IT 2006 and 2009 and JP 2014, the 5LL for IT 2007-2011 and JP 2010, 2013-2014, and the 5LSt for IT 2007-2011.

We have computed the KS, CM, and AD statistics of the out-of-sample data sets when evaluated on the in-sample estimated distributions, and also the corresponding log-likelihoods, AIC, BIC, and HQC information criteria. It is known that this procedure shows more parsimony in the selected models than the full sample estimations.

\begin{table}[htbp]
\centering {
    \begin{tabular}{lrrrrrrr}
    \hline
         & \multicolumn{1}{l}{Dist. Eq.}& \multicolumn{1}{l}{$k$} & \multicolumn{1}{l}{Min KS} &       & \multicolumn{1}{l}{Min CM} &       & \multicolumn{1}{l}{Min AD} \\
    LN    & (\ref{modelLN})& 2 & 0     &       & 0     &       & 0 \\
    DPLN  & (\ref{modelDPLN})& 4 & 0     &       & 0     &       & 1 \\
    GB2   & (\ref{modelGB2})& 4 & 0     &       & 0     &       & 0 \\
    LNSNP & (\ref{modelLNSNP})& 6 & 2     &       & 2     &       & 2 \\
    2LN   & (\ref{model2LN})& 5 & 1     &       & 2     &       & 0 \\
    3LN   & (\ref{model3LN})& 8 & 5     &       & 5     &       & 6 \\
    4LN   & (\ref{model4LN})& 11 & 2     &       & 5     &       & 7 \\
    5LN   & (\ref{model5LN})& 14 & 1     &       & 4     &       & 4 \\
    2LL   & (\ref{model2LL})& 5 & 6     &       & 3     &       & 2 \\
    3LL   & (\ref{model3LL})& 8 & 4     &       & 2     &       & 1 \\
    4LL   & (\ref{model4LL})& 11 & 7     &       & 2     &       & 5 \\
    5LL   & (\ref{model5LL})& 14  & 5     &       & 5     &       & 4 \\
    2LSt12 & (\ref{model2LSt12})& 5 & 4     &       & 9     &       & 2 \\
    2LSt39 & (\ref{model2LSt39})& 5 & 3     &       & 1     &       & 0 \\
    3LSt  & (\ref{model3LSt})& 8 & 4     &       & 5     &       & 7 \\
    4LSt  & (\ref{model4LSt})& 11 & 4     &       & 4     &       & 6 \\
    5LSt  & (\ref{model5LSt})& 14 & 3     &       & 2     &       & 4 \\
    Total & & & 51    &       & 51    &       & 51 \\
 \hline
\end{tabular}%
}
\caption{Summary of the results of the minimum KS, CM, AD tests' statistics for each model and the out-of-sample data into the in-sample estimations.}
\label{KSCMADisoos}
\end{table}

We can see in Table~\ref{KSCMADisoos} that there is not a clear ranking in the classification of the different distributions with regards to the outcome of KS, CM, AD statistics when considering
non-mixture or mixture distributions, and amongst these last ones, there is not a clear ranking between a lower number of components and higher number of components, as the lowest statistics are scattered across different distributions.

As with regards to the information criteria, we show in Table~\ref{AICBICHQCisoos} the corresponding summary. We see that the top-five minimum information criteria are obtained for the 3LN (43 times out of 153), the 2LN (21), the 2LL (20), the 4LSt (16), and the 4LN (13). Thus we observe that none of the non-mixture models is amongst the top minimum information criteria distributions and that amongst the mixture models, there is a majority of minimum information criteria with only three (3LN) or two (2LN, 2LL) components, shown in this way parsimony in the number of parameters of the descriptive power of
out-of-samples of different distributions.

\begin{table}[htbp]
\centering {
    \begin{tabular}{lrrrrrrr}
\hline
       & \multicolumn{1}{l}{Dist. Eq.}& \multicolumn{1}{l}{$k$}   & \multicolumn{1}{l}{Sel AIC} &       & \multicolumn{1}{l}{Sel BIC} &       & \multicolumn{1}{l}{Sel HQC} \\
    LN    & (\ref{modelLN}) & 2& 0     &       & 0     &       & 0 \\
    DPLN  & (\ref{modelDPLN})& 4 & 1     &       & 4     &       & 2 \\
    GB2   & (\ref{modelGB2})& 4 & 0     &       & 4     &       & 0 \\
    LNSNP & (\ref{modelLNSNP})& 6 & 1     &       & 3     &       & 3 \\
    2LN   & (\ref{model2LN})& 5 & 3     &       & 13    &       & 5 \\
    3LN   & (\ref{model3LN})& 8 & 17    &       & 6     &       & 20 \\
    4LN   & (\ref{model4LN})& 11 & 7     &       & 2     &       & 4 \\
    5LN   & (\ref{model5LN})& 14 & 1     &       & 0     &       & 0 \\
    2LL   & (\ref{model2LL})& 5& 1     &       & 14    &       & 5 \\
    3LL   & (\ref{model3LL})& 8& 1     &       & 1     &       & 2 \\
    4LL   & (\ref{model4LL})& 11& 3     &       & 1     &       & 3 \\
    5LL   & (\ref{model5LL})& 14& 1     &       & 0     &       & 1 \\
    2LSt12 & (\ref{model2LSt12})& 5& 0     &       & 1     &       & 0 \\
    2LSt39 & (\ref{model2LSt39})& 5 & 0     &       & 2     &       & 0 \\
    3LSt   & (\ref{model3LSt})& 8& 1     &       & 0     &       & 1 \\
    4LSt  & (\ref{model4LSt})& 11& 11    &       & 0     &       & 5 \\
    5LSt  & (\ref{model5LSt})& 14& 3     &       & 0     &       & 0 \\
    Total & & & 51    &       & 51    &       & 51 \\
 \hline
\end{tabular}%
}
\caption{Summary of the results of the selection of the distributions for the in-sample out-of-sample analysis by information criteria AIC, BIC, HQC.}
\label{AICBICHQCisoos}
\end{table}



\subsection{Doubly truncated data sets and distributions}\label{ttsamplesresults}

In this subsection, we will consider the doubly truncated data sets and distributions. That is, we drop from the full data sets the lowest 10\% of observations and the top 0.1\% of the observations, as is suggested for example in \cite{Fio20}.
Also, the distributions are modified as follows \citep{JohKotBal94}.
Suppose we have a random variable $X$ that is distributed according to some PDF $f(x)$, with
CDF ${\mathrm{cdf}}(x)$, both of which have support equal to $(0,\infty)$.
Suppose we wish to know the PDF of the random variable after restricting the support to $[a,b]$ with $0<a<b<\infty$.
Then
\begin{equation}
f(x|a\leq X\leq b)={\frac {g(x)}{{\mathrm{cdf}}(b)-{\mathrm{cdf}}(a)}}
\label{trunc}
\end{equation}
where $g(x)=f(x)I(\{a\leq x\leq b\})$ and $I$ is the indicator function. If the support of the non-truncated distribution is $(-\infty,\infty)$, the procedure is entirely similar. The formulae for the CDFs of our considered distributions are available upon request.
For the non-mixture models we perform this truncation as just explained. For the mixture models we perform the truncation of each of the components first, and after that taking the mixture of them. There is the alternative of taking first the mixture of the components and after that, doubly truncating the so obtained mixture. We feel that the first alternative provides better results as the estimations are more stable and more of them out of the possible ones are obtained.
In all cases, the doubly truncated distributions corresponding to the ones described in Section~\ref{distros} will be appended with \lq\lq tt\rq\rq.

\begin{table}[htbp]
\centering {
   \begin{tabular}{lrrrr}
   \hline
          & \multicolumn{1}{l}{$k$} & \multicolumn{1}{l}{Min KS} & \multicolumn{1}{l}{Min CM} & \multicolumn{1}{l}{Min AD} \\
    LNtt  & 2 & 0     & 0     & 0 \\
    DPLNtt & 4 & 0     & 0     & 0 \\
    GB2tt & 4 & 0     & 0     & 0 \\
    LNSNPtt & 6 & 0     & 0     & 0 \\
    2LNtt & 5 & 0     & 0     & 0 \\
    3LNtt & 8& 3     & 0     & 0 \\
    4LNtt & 11 & 1     & 0     & 1 \\
    5LNtt & 14 & 5     & 1     & 5 \\
    2LLtt & 5 & 0     & 0     & 0 \\
    3LLtt & 8 & 4     & 3     & 1 \\
    4LLtt & 11 & 7     & 11     & 7 \\
    5LLtt & 14 & 12    & 27    & 24 \\
    2LSt12tt & 5 & 2     & 0     & 0 \\
    2LSt39tt & 5 & 0     & 0     & 0 \\
    3LSttt & 8 & 4     & 0     & 0 \\
    4LSttt & 11 & 5     & 3    & 1 \\
    5LSttt & 14 & 8     & 6    & 12 \\
    Total & & 51    & 51    & 51 \\
 \hline
\end{tabular}%
}
\caption{Summary of the results of the minimum KS, CM, AD tests' statistics for each \lq\lq tt\rq\rq\ model and the doubly truncated data sets.}
\label{KSCMADtt}
\end{table}

We observe in Table~\ref{KSCMADtt} that the top-four lowest statistics correspond to the 5LLtt (63 times out of 153), 5LSttt (26), 4LLtt (25), and 5LNtt (11) and that the non-mixture distributions never provide the lowest statistics.
Thus there is a tendency to have lower statistics with five-component mixture distributions (5LLtt and 5LSttt) followed by a four-component one (4LLtt), although next is also a five-component one (5LNtt). This is not unexpected for the reasons we have pointed out before, namely that the goodness-of-fit does not penalize the higher number of parameters.

To consider that, let us have a look at the information criteria. We show in Table~\ref{AICBICHQCtt} the summary of the results.
We see that the top-five most selected distributions are the 5LSttt (28 times out of 153), 3LNtt (14), 2LLtt (12), 5LLtt (11), and 3LSttt (8). The non-mixture distributions are selected 5 times in total out of 153, but never the LNtt. Thus we do not
obtain a clear parsimony in this case with the number of the parameters, as there is an ample selection of five-component mixture distributions. This may be caused by the difficulty of modelling the upper tail of the distributions, even after having removed the top 0.1\% of the observations, in line with \cite{Ram22} and references therein.

\begin{table}[htbp]
\centering {
    \begin{tabular}{lrrrr}
\hline
          & \multicolumn{1}{l}{$k$}& \multicolumn{1}{l}{Min AIC} & \multicolumn{1}{l}{Min BIC} & \multicolumn{1}{l}{Min HQC} \\
    LNtt & 2 & 0     & 0     & 0 \\
    DPLNtt & 4 & 0     & 0     & 1 \\
    GB2tt & 4 & 0     & 0     & 2 \\
    LNSNPtt & 6 & 0     & 0     & 2 \\
    2LNtt & 5 & 0     & 0     & 2 \\
    3LNtt & 8 & 7     & 0     & 7 \\
    4LNtt & 11 & 3     & 0     & 3 \\
    5LNtt & 14 & 0     & 0     & 0 \\
    2LLtt & 5 & 2     & 0    & 10 \\
    3LLtt & 8 & 3     & 0     & 2 \\
    4LLtt & 11 & 2     & 0     & 1 \\
    5LLtt & 14 & 8     & 0     & 3 \\
    2LSt12tt & 5 & 2     & 0     & 2 \\
    2LSt39tt & 5 & 0     & 0     & 2 \\
    3LSttt & 8 & 5    & 0     & 3 \\
    4LSttt & 11 & 2     & 0     & 0 \\
    5LSttt & 14 & 17     & 0     & 11 \\
    Total & & 51    & 51    & 51 \\
 \hline
\end{tabular}%
}
\caption{Summary of the results of the selection of the distributions for the doubly truncated data sets and distributions by information criteria AIC, BIC, HQC.}
\label{AICBICHQCtt}
\end{table}



\subsection{Fokker--Planck equations for the preferred models}\label{FPeqs}

We will briefly show the solution for $b(y,t)$ of (\ref{solb}) associated to the Fokker--Planck equation (\ref{fpeq}) in the case of the 4LSt (full samples) and 5LSttt (doubly truncated samples). The corresponding preferred model for the in-sample out-of-sample analysis, namely the 3LN, can be treated in a similar way and it is already presented in \cite{PenPueRamSan22} (with a variable $g$ instead of $y$, it is easy to perform the substitution).

Let us begin with the 4LSt. We denote first the natural logarithm of $x>0$ by $y=\ln(x)\in(-\infty,\infty)$.
The PDF of the ordinary Student's $t$ distribution in the variable $y$ is given by
$$
f_{{\rm St}}(y;\mu,\sigma,\nu)=
\frac{\Gamma\left(\frac{\nu+1}{2}
\right)}{\Gamma\left(\frac{\nu}{2}\right)\sqrt{\pi \nu}\sigma}
\left(1+\frac{1}{\nu}\left(\frac{y-\mu}{\sigma}
\right)^2
\right)^{-\frac{\nu+1}{2}}\nonumber
$$
and the corresponding CDF is given by
$$
{\rm cdf}_{{\rm St}}(y;\mu,\sigma,\nu)=
\frac{1}{2}+
\frac{\Gamma\left(\frac{1+\nu}{2}\right)}
{\Gamma\left(\frac{\nu}{2}\right)\sqrt{\pi\nu}\sigma}
(y-\mu)\,{}_2F_{1}\left(\frac{1}{2},\frac{1+\nu}{2},
\frac{3}{2},-\frac{(y-\mu)^2}{\nu \sigma^2}
\right)
$$
where ${}_2F_{1}$ denotes the Gaussian or ordinary hypergeometric function.
Let us denote for the sake of brevity the corresponding time-dependent density function by
\begin{eqnarray}
j_{\rm 4St}(y,t)
&=&p_1(t)f_{{\rm St}}(y;\mu_1(t),\sigma_1(t),4)\nonumber\\
& &+p_2(t)f_{{\rm St}}(y;\mu_2(t),\sigma_2(t),12)\nonumber\\
& &+p_3(t)f_{{\rm St}}(y;\mu_3(t),\sigma_3(t),39)\nonumber\\
& &+(1-p_1(t)-p_2(t)-p_3(t))f_{{\rm St}}(y;\mu_4(t),\sigma_4(t),100)\nonumber
\end{eqnarray}

Also, let us denote for convenience the following expression:
$$
k(y;\mu,\sigma,\nu,s)=\dot{\mu}
+(y-\mu)\frac{\dot{\sigma}}{\sigma}
-\frac{s^2(1+\nu)(y-\mu)}{2((y-\mu)^2+\nu \sigma^2)}
$$
where $\mu\in\mathbb{R}$ and $\sigma,\nu>0$.
$\mu$ and $\sigma$ are supposed to depend smoothly on $t$ (explicit dependence is omitted for notational simplicity) and $\nu$ is fixed (\emph{a priori}), the dot means derivative with respect to $t$, and $s>0$ is a real constant. In this expression, taking $\nu\rightarrow\infty$ yields the corresponding expression for the normal distribution, as it should be, presented in \cite{PenPueRamSan22} (for the variable $g$ instead of $y$).
Let us define also the quantities
\begin{eqnarray}
&&\pi_1(y,t)=\frac{{\rm cdf}_{{\rm St}}(y;\mu_1,\sigma_1,4)-{\rm cdf}_{{\rm St}}(y;\mu_4,\sigma_4,100)}
{j_{\rm 4St}(y,t)}\nonumber\\
&&\pi_2(y,t)=\frac{{\rm cdf}_{{\rm St}}(y;\mu_2,\sigma_2,12)-{\rm cdf}_{{\rm St}}(y;\mu_4,\sigma_4,100)}
{j_{\rm 4St}(y,t)}\nonumber\\
&&\pi_3(y,t)=\frac{{\rm cdf}_{{\rm St}}(y;\mu_3,\sigma_3,39)-{\rm cdf}_{{\rm St}}(y;\mu_4,\sigma_4,100)}
{j_{\rm 4St}(y,t)}\nonumber
\end{eqnarray}
and the time-dependent posterior probabilities
\begin{eqnarray}
&&\tau _{1}(y,t)=p_{1}
f_{{\rm St}}(y;\mu_{1},\sigma_{1},4)/j_{\rm{4St}}(y,t)
\nonumber\\
&&\tau _{2}(y,t)=p_{2}
f_{{\rm St}}(y;\mu_{2},\sigma_{2},12)/j_{\rm{4St}}(y,t)
\nonumber\\
&&\tau _{3}(y,t)=p_{3}
f_{{\rm St}}(y;\mu_{3},\sigma_{3},39)/j_{\rm{4St}}(y,t)
\nonumber\\
&&\tau _{4}(y,t)=(1-p_{1}-p_{2}-p_{3})
f_{{\rm St}}(y;\mu_{4},\sigma_{4},100)/j_{\rm{4St}}(y,t)
\nonumber
\end{eqnarray}%
Then, if we select $a(y,t)=s^2$, with $s>0$ being a real \emph{constant}, the $b(y,t)$ given by (\ref{solb}) turns into
\begin{eqnarray}
b(y,t)&=&k(y;\mu_1,\sigma_1,4,s)\tau_1(y,t)+
k(y;\mu_2,\sigma_2,12,s)\tau_2(y,t)\nonumber\\
& &+
k(y;\mu_3,\sigma_3,39,s)\tau_3(y,t)+
k(y;\mu_4,\sigma_4,100,s)\tau_4(y,t)\nonumber\\
& &-\dot{p}_1\pi_1(y,t)
-\dot{p}_2\pi_2(y,t)-\dot{p}_3\pi_3(y,t)\nonumber
\label{b4St}
\end{eqnarray}%
so that we obtain that $f(y,t)=j_{\rm 4St}(y,t)$ is in this case a solution of the corresponding Fokker--Planck equation (\ref{fpeq}). The sign of the drift term is indefinite on this occasion.
This is related to the results of  \cite{CamRam21,PenPueRamSan22}.

For the doubly truncated 5LSt (5LSttt) case, let us denote again first $y=\ln(x)$.
Then, we doubly truncate the corresponding Student's $t$ distribution to give ($y_{\rm min}=\ln(x_{\rm min})$ and $y_{\rm max}=\ln(x_{\rm max})$)
$$
f(y;y_{\rm min},y_{\rm max},\mu,\sigma,\nu)=
\frac{f_{{\rm St}}(y;\mu,\sigma,\nu)
I(\{y_{\rm min}\leq y\leq y_{\rm max}\})}{{\rm cdf}_{{\rm St}}(y_{\rm max};\mu,\sigma,\nu)-{\rm cdf}_{{\rm St}}(y_{\rm min};\mu,\sigma,\nu)}
$$
with support $-\infty<y_{\rm min}\leq y\leq y_{\rm max}<\infty$ and the corresponding CDF
being given by
$$
{\rm cdf}(y;y_{\rm min},y_{\rm max},\mu,\sigma,\nu)=
\frac{({\rm cdf}_{{\rm St}}(y;\mu,\sigma,\nu)-{\rm cdf}_{{\rm St}}(y_{\rm min};\mu,\sigma,\nu))I(\{y_{\rm min}\leq y\leq y_{\rm max}\})}
{{\rm cdf}_{{\rm St}}(y_{\rm max};\mu,\sigma,\nu)-{\rm cdf}_{{\rm St}}(y_{\rm min};\mu,\sigma,\nu)}
$$
The corresponding time-dependent mixture is then
\begin{eqnarray}
j_{\rm{5Sttt}}(y,t)&=&
p_1(t)f(y;y_{\rm min}(t),y_{\rm max}(t),\mu_1(t),\sigma_1(t),4)\nonumber\\
& &+p_2(t)f(y;y_{\rm min}(t),y_{\rm max}(t),\mu_2(t),\sigma_2(t),12)\nonumber\\
& &+p_3(t)f(y;y_{\rm min}(t),y_{\rm max}(t),\mu_3(t),\sigma_3(t),39)\nonumber\\
& &+p_4(t)f(y;y_{\rm min}(t),y_{\rm max}(t),\mu_4(t),\sigma_4(t),100)\nonumber\\
& &+(1-p_1(t)-p_2(t)-p_3(t)-p_4(t))f(y;y_{\rm min}(t),y_{\rm max}(t),\mu_5(t),\sigma_5(t),200)\nonumber
\end{eqnarray}

Then, let us denote again for convenience the following expression:
$$
k(y;y_{\rm min},y_{\rm max},\mu,\sigma,\nu,s)=
\frac{s^2}{2f}\frac{\partial f}{\partial y}
-\frac{1}{f}\left(\dot{\sigma}\frac{\partial{\rm cdf}}{\partial \sigma}+\dot{\mu}\frac{\partial{\rm cdf}}{\partial \mu}+\dot{y}_{\rm max}\frac{\partial{\rm cdf}}{\partial y_{\rm max}}+\dot{y}_{\rm min}\frac{\partial{\rm cdf}}{\partial y_{\rm min}}\right)
$$
where $\mu,y_{\rm min},y_{\rm max}\in\mathbb{R}$, $\sigma>0$, and are supposed to depend smoothly on $t$ (explicit dependence is omitted for notational simplicity), and $\nu>0$ is kept fixed; the dot means derivative with respect to $t$, and $s>0$ is a real constant.
Let us define also the quantities
\begin{eqnarray}
&&\pi_1(y,t)=\frac{{\rm cdf}(y;y_{\rm min},y_{\rm max},\mu_1,\sigma_1,4)
-{\rm cdf}(y;y_{\rm min},y_{\rm max},\mu_5,\sigma_5,200)}
{j_{\rm{5Sttt}}(y,t)}\nonumber\\
&&\pi_2(y,t)=\frac{{\rm cdf}(y;y_{\rm min},y_{\rm max},\mu_2,\sigma_2,12)
-{\rm cdf}(y;y_{\rm min},y_{\rm max},\mu_5,\sigma_5,200)}
{j_{\rm{5Sttt}}(y,t)}\nonumber\\
&&\pi_3(y,t)=\frac{{\rm cdf}(y;y_{\rm min},y_{\rm max},\mu_3,\sigma_3,39)
-{\rm cdf}(y;y_{\rm min},y_{\rm max},\mu_5,\sigma_5,200)}
{j_{\rm{5Sttt}}(y,t)}\nonumber\\
&&\pi_4(y,t)=\frac{{\rm cdf}(y;y_{\rm min},y_{\rm max},\mu_4,\sigma_4,100)
-{\rm cdf}(y;y_{\rm min},y_{\rm max},\mu_5,\sigma_5,200)}
{j_{\rm{5Sttt}}(y,t)}\nonumber
\end{eqnarray}
and the time-dependent posterior probabilities
\begin{eqnarray}
&&\tau _{1}(y,t)=p_{1}f(y;y_{\rm min},y_{\rm max},\mu_1,\sigma_1,4)/j_{\rm{5Sttt}}(y,t)\nonumber\\
&&\tau _{2}(y,t)=p_{2}f(y;y_{\rm min},y_{\rm max},\mu_2,\sigma_2,12)/j_{\rm{5Sttt}}(y,t)\nonumber\\
&&\tau _{3}(y,t)=p_{3}f(y;y_{\rm min},y_{\rm max},\mu_3,\sigma_3,39)/j_{\rm{5Sttt}}(y,t)\nonumber\\
&&\tau _{4}(y,t)=p_{4}f(y;y_{\rm min},y_{\rm max},\mu_4,\sigma_4,100)/j_{\rm{5Sttt}}(y,t)\nonumber\\
&&\tau _{5}(y,t)=(1-p_{1}-p_{2}-p_{3}-p_{4})f(y;y_{\rm min},y_{\rm max},\mu_5,\sigma_5,200)
/j_{\rm{5Sttt}}(y,t)\nonumber
\end{eqnarray}%
Then, if we select $a(y,t)=s^2$, being again $s>0$ a constant, the $b(y,t)$ given by (\ref{solb}) turns into
\begin{eqnarray}
b(y,t)&=&
k(y;y_{\rm min},y_{\rm max},\mu_1,\sigma_1,4,s)\tau_1(y,t)\nonumber\\
& &+k(y;y_{\rm min},y_{\rm max},\mu_2,\sigma_2,12,s)\tau_2(y,t)\nonumber\\
& &+k(y;y_{\rm min},y_{\rm max},\mu_3,\sigma_3,39,s)\tau_3(y,t)\nonumber\\
& &+k(y;y_{\rm min},y_{\rm max},\mu_4,\sigma_4,100,s)\tau_4(y,t)\nonumber\\
& &+k(y;y_{\rm min},y_{\rm max},\mu_5,\sigma_5,200,s)\tau_5(y,t)\nonumber\\
& &-\dot{p}_1\pi_1(y,t)
-\dot{p}_2\pi_2(y,t)
-\dot{p}_3\pi_3(y,t)
-\dot{p}_4\pi_4(y,t)\nonumber
\label{b5Sttt}
\end{eqnarray}%
so that we obtain that $f(y,t)=j_{\rm{5Sttt}}(y,t)$ is in this case a solution of the corresponding Fokker--Planck equation (\ref{fpeq}). The sign of the drift term is indefinite also on this occasion.
This development for mixtures of doubly truncated distributions is new as far as we know. Also, it is easy to substitute the Student's $t$ in these mixtures by another distribution to give the corresponding expressions for other mixtures defined on $(-\infty,\infty)$ or doubly truncated ones.

These solutions of the Fokker--Plank equation for the preferred time-dependent density functions might lead to a new way of simulating and/or perhaps forecasting the firm's sales of the different countries, leaving that research to a subsequent paper.

\section{Conclusions}

We have considered ORBIS data sets for six different countries, on ample time intervals, and a total of 51 samples, so we believe that the results are quite robust.

We have studied the fit of 17 different statistical models obtained from the classic and recent literature on sizes distributions in the social sciences and Economics, and applied them to the firm size distribution measured by sales, with mixtures of distributions or not. Out of them, there is a classical model, the lognormal (LN), that is strongly not selected always for all samples. When the full data sets are considered, the most often selected distribution is the 4LSt with 11 parameters, and it is not a distribution with the highest number of parameters out of that considered (for example, the 5LN, 5LL, and 5LSt have all 14 parameters). When performing in-sample out-of-sample analysis, the descriptive power of the 3LN has the best results out of our 17 distributions, and again the LN is never selected.
Finally, when we doubly truncate the data and the distributions, the 5LSttt is the most often selected distribution, and never the LNtt.
This may be due to the that there may be firms
which operate by different mechanisms and then conform to some sub-populations in each of the samples, as in other contexts \citep{BelCleBul14,Kun20} and as described in \cite{McLPee00}.
Depending on the method of measurement, for the best-fit distribution, it is necessary to
consider a mixture of at least three distributions or even four or five ones.

Thus we have two clear results: First, in none of the situations the lognormal or variations thereof is the selected distribution, and depending on the situation one mixture distribution
\citep{KwoNad19,BanChiPrePueRam19,GuaTos19,GuaTos19b,Su19,PueRamSan20,PueRamSanArr20}
 is preferred to other ones, when describing firm size distribution measured by sales.
Second, it is clearly stated the difficulty of modelling the tails of the full or doubly truncated data sets: A higher number of components in the mixture should be taken into account to obtain better information criteria for these samples.

Also, the mixtures considered in this paper, when taken as time-dependent ones, can be shown to be solutions of a Fokker--Planck equation with constant diffusion term as described in the introduction of this paper, following the lines of \cite{CamRam21,PenPueRamSan22}. We defer the use of such newly introduced solutions in simulating or forecasting to another paper.

\section*{Author contributions}
Shouji Fujimoto: Conceptualization, data curation, formal analysis, funding acquisition, investigation, methodology, software, supervision, validation, visualization, writing-original draft.
Atushi Ishikawa: Conceptualization, data curation, formal analysis, funding acquisition, investigation, methodology, software, supervision, validation, visualization, writing-original draft.
Till Massing: Conceptualization, data curation, formal analysis, funding acquisition, investigation, methodology, software, supervision, validation, visualization, writing-original draft. Takayuki Mizuno: Conceptualization, data curation, formal analysis, funding acquisition, investigation, methodology, software, supervision, validation, visualization, writing-original draft.
Arturo Ramos: Conceptualization, data curation, formal analysis, funding acquisition, investigation, methodology, resources, software, validation, visualization, writing-original draft.

\section*{Competing interests statement}
The authors declare to have no competing interests concerning the research carried out in this article.

\section*{Data availability statement}
The Edition of 2016 ORBIS database used in this study has been obtained upon payment from Bureau van Dijk, and we have signed a confidentiality agreement so we cannot disclose the data sets. For the sake of replication, interested researchers can buy the same edition of such database.

\section*{Acknowledgments}

We thank Profs. Josefina and Rafael Cabeza-Laguna for their help in the computations for the estimation of the doubly truncated mixture 5LSttt.

\bibliographystyle{apalike}
\bibliography{biblio}

\end{document}